**Intelligent and distributed production control**

André Thomas[1], Damien Trentesaux[2*], Paul Valckenaers[3]

[1]Research Centre for Automatic Control (CRAN),CNRS (UMR 7029), Nancy University, 27, rue du merle blanc, 88000 Epinal, France

Andre.Thomas@cran.uhp-nancy.fr

[2]Université Lille Nord de France, F-59000 Lille, France

UVHC, TEMPO-Lab., Le mont houy, F-59313 Valenciennes, cedex 9, France

Damien.Trentesaux@univ-valenciennes.fr

[3]K.U. Leuven - Mechanical engineering Celestijnenlaan 300, B-3001 Leuven, Belgium

Paul.Valckenaers@mech.kuleuven.be

*Corresponding author

## Abstract

This editorial introduces the special issue of the Springer journal, Journal of Intelligent manufacturing, on Intelligent and distributed production control. This special issue contains selected papers of the 13th IFAC Symposium on Information Control Problems in Manufacturing (Bakhtadze and Dolgui, 2009) – INCOM'2009. The papers presented in this special issue have been selected according to their high quality and according to their specific way of addressing the variety of issues dealing with intelligent and distributed production control. Preceding global discussion on the state of the art in intelligent and distributed production control is provided as well as a tentative guideline for future works in this area.

## Introduction

The Springer journal, Journal of Intelligent manufacturing, welcomes a special issue ( SI) on Intelligent and distributed production control. This SI contains selected papers of the 13th IFAC Symposium on Information Control Problems in Manufacturing (Bakhtadze and Dolgui, 2009) – INCOM'2009. Taking place from June 3rd to 5th, 2009 in Moscow (Russia), the symposium was organized by the V.A. Trapeznikov Institute of Control Sciences of the Russian Academy of Sciences and sponsored by IFAC, IFIP, Russian Academy of Sciences and GDR MACS/CNRS (French National Council for Scientific Research). It has been a long time that Intelligent and distributed issues in production control are addressed by the international community, and especially by the IFAC and this conference gave us the opportunity to propose to the chairs of INCOM a special issue devoted to this topic. This *edito* is organized as follows: first, intelligent and distributed production control is introduced as a key concept for future researches. A set of major issues are then presented and discussed. The research articles of this SI are introduced as possible ways to address some of these issues. This *edito* concludes with a set of short term and long term prospects.

## Stakes and current issues in intelligent and distributed production control

Historically, 'centralized' approaches (based upon the federative concept of CIM – Computer Integrated Manufacturing) have been implemented thanks to MRP2 (Manufacturing Resources Planning) systems and more recently, to ERP (Enterprise Resources Planning) systems, with tools and methods mainly based on operational research concerning production activity control. In centralized approaches, decision making is hierarchically broadcasted from the higher decisional levels down to the operational units. The success of these approaches mainly holds in their ability to provide long term optimization of production planning and scheduling given a relatively stable industrial context.

Facing the eighties' market challenges other decision making philosophies and strategies have emerged. Requirements for more and more reactivity and flexibility have led to the implementation of first 'distributed' approaches such as anthropocentric and visual management methods (kanban, operators' empowerment, etc.). Unfortunately, these new ways to pilot and control the material flows have led to 'black boxes' in management systems, and have highlighted the need for more and more real-time closed-loop information systems. It has been shown (Klein, 2008) that in such kanban systems, very short term priority management is always a key issue. More recently, in the 90's, Production and Supply Chain Systems have changed from the traditional mass production led by products to the mass customization in order to face the increase of the global market competition. High competition between enterprises and market volatility led then enterprises to be more agile (Christopher, 1992). Agility, from the point of view of production control, may be seen as the ability to operate with a high level of coordination and proactivity throughout the supply-chain, and at the same time to react efficiently to disturbances on the shop floor while taking into account the increasing process complexity (variabilities, high product variety, reconfiguration issues).

Gradually and following information technology improvements, it seemed obvious that to give to the physical system entities (parts, resources, ...) some decision making capabilities could be a new way to face this ever unsolved issue. One argument was that, in 'centralized' approaches, the time spent to inform the correct controller within the hierarchy (bottom-up), and then to decide and to apply the decision (top-down) generates lags and instabilities. Another argument was these approaches, despite their ability to provide near-optimal behavior within fully static and deterministic environment, could not easily face disturbances and could not evolve easily with of the environment. The constantly increasing power of the central calculator hardly handles the induced complexity; even if ERP systems are now widespread, these systems does not fully satisfy industrial needs always seeking for more agility. A recent study highlighted that, for European factory equipment suppliers, the priority among 10 major concerns was the need for "intelligent products", including self-optimizing systems [Schreiber (2007) cited in Sauer (2008)]. Industrial requirements have clearly evolved from the usual traditional performance criteria, described in terms of static optimality or near-optimality, towards new performance criteria, described in terms of reactivity, adaptability and robustness. A growing number of industrialists now want control systems that provide satisfactory, adaptable and robust solutions rather than optimal solutions that require meeting several hard assumptions.

Consequently, since 90's, an increasing research activity in manufacturing systems control has moved from traditional centralized approaches to distributed architectures. Among other, fully heterarchical architectures promote production control by distributing every decision capacities in autonomous entities, without any centralised view of the shop floor elements status. In order to ensure consistency of decision making, more pragmatic approaches based on hybrid control combining the predictability of the centralized control with the agility and robustness against disturbances of the heterarchical control have been designed, enabling the integration of optimization models with cooperation and local autonomy features. To name a few, the concepts of Holonic Manufacturing Systems (HMS) (Babiceanu and Chen, 2006), of Product Driven Systems (PDS), of Intelligent Manufacturing Systems (IMS), and of Agent-Based Manufacturing (Maturana, 1999) have been proposed to design these future manufacturing systems. These concepts advocate that the products, and more globally, all the production resources can be modeled as an association between two parts, a physical part and an informational one. These intelligent systems shall assist operators and managers in manufacturing and supply chain control.

The common denominator for all these approaches is to bring intelligence and autonomy as near as possible to (or even in) the physical system. The idea was to permit the decisional entities to work together so as to react quickly in an autonomous way, within constraints, instead of requesting control decisions from upper decisional levels, which was generating response time lags. In these approaches, interaction processes other than coordination appear, mainly, negotiation and cooperation (Mařík and Lazansky, 2007). However, negotiation and cooperation led to new problems, for example, the need to prove deadlock avoidance mechanisms and more generally, the need to prove that sufficient level of performance can be attained. As a consequence, the maturity level of these approaches is still low, hardening of even forbidding real implementations. Globally, four challenges are to be addressed to reach a sufficient maturity level (Trentesaux, 2009): performance guarantees (how to prove the abilities of the intelligent and distributed control ?), emergence engineering (how to control the emerging behavior in a desired way?), interoperability and norms (how the intelligent and distributed control can be integrated and interfaced with existing information systems ?), and last, the development, scalability and costs (how can an industrial can implement an intelligent and distributed control system and measure his return-on-investment compared to classical centralized solutions ?).

Despite this lack of maturity, the research and the industrial interest in this area are currently boosted by several new technologies and software development that are appearing, offering new abilities potentially useful to increase flexibility and reactivity. Facing these new trends, a lot of new research works are focusing on identification technologies, like electronic (Auto ID) or biometric (vision) ones. Radio Frequency Identification technology (RFID) represents a quick and safe way to track products, opening the way of linking informational and physical flows which still remain an important research challenge (Plossl, 1993), and providing an accurate, real time vision of the shop floor. More and more efforts are also put in the development of norms enabling the implementation of distributed and intelligent production control (eg., International Industrial Standard IEC61499, IEEE Foundations of Intelligent Physical Agents - FIPA Standards Committee). These new technologies appear like a catalyst to change the fifty years old way of controlling production through traditional MRP2 systems (Thomas, 2008).

# The contributions

In this SI, 11 papers have been selected after a strong peer review process implying reviewers from the INCOM conference to evaluate the improvements of the papers and new reviewers from the *Journal of Intelligent Manufacturing* community. These papers are shortly introduced in the following:

Aissani N., Bekrar A., Trentesaux D. and Beldjilali B. propose a model for adaptive scheduling in multi-site companies. A multi-agent approach is adopted in which intelligent agents have reactive learning capabilities based on reinforcement learning. Experimentations and simulations inspired from a real case study demonstrate the applicability and the effectiveness of the model in terms of both optimality and reactivity.

In Khalgui M., Mosbahi O., Hanisch H-M., Li Z., authors define an architecture of reconfigurable multi-agent systems in which a Reconfiguration Agent is affected to each device of the execution environment to apply local reconfigurations, and a Coordination Agent is proposed for coordinations between devices in order to guarantee safe and adequate distributed reconfigurations.

The paper of Leitão P., Mendes J. M., Bepperling A., Cachapa D., Colombo A. W. and Restivo F. discusses the integration of 2D/3D digital software tools with Petri net based service-oriented frameworks to allow the design, configuration, analysis, validation, simulation, monitoring and control of manufacturing systems in a virtual environment and its posterior smooth migration into the real "physical" environment.

Ostrosi E., Fougères A.-J., Ferney M. and Klein D. propose a Fuzzy Configuration Grammar based agents to assist collaborative and distributed design for product configuration. Based on the distributed fuzzy models, fuzziness of interactions during the collaborative and distributed design for configuration, a computational approach for product configuration is developed.

The objective of the paper of Pannequin R. and Thomas A., is to propose another interpretation of stigmergy in such IMS context. In this paper, the authors propose to come back to the basics, that is to say that information (pheromone) is attached to the products. Agent oriented components which implement stigmergic design pattern are presented, are firstly applied to a laboratory platform, and secondly on an industrial test-case.

In their paper, Radakovič M., Obitko M. and Mařík V. discuss the necessity of explicit definition of both declarative and procedural knowledge and propose explicit procedural knowledge handling. Sharing and distribution of such knowledge is discussed and illustrated on an implemented transportation system example. They also introduce the utilization of discussed architecture for explicit specification of agent behavior in failures patterns handling and smart grid configuration scenario.

To elaborate daily employees' assignment to workstations in a workshop and to propose a production planning, Sabar M., Montreuil B. and Frayret J.-M. propose a multi-agent based algorithm for personnel scheduling and rescheduling in a dynamic environment of a paced multi-product assembly center.

The paper of Saint Germain B., Valckenaers P., Van Belle J., Verstraete P. and Van Brussel H. presents a decision-making pattern that uses trust mechanisms based on past performance. It infers how information is to be understood and the uncertainty on the expected behavior. This pattern makes a contribution by enabling cooperation in large systems of systems without requiring conventional integration.

The main concern of the research work of Tounsi J., Habchi G., Boissière J. and Azaiez S. is to analyse and model supply chains in the particular context of small and medium enterprises in the field of mechatronic. The first contribution is to propose a generic metamodel for supply chains and the second one lies in the formalisation of the dynamic behaviour of the concepts in the metamodel.

In a self-moving products context, the tracking of them is a major challenge. The paper of Véjar A. and Charpentier P. presents a general framework in location and manufacturing applications with objective to reproduce the manufacturing system dynamics in an adaptive simulation scheme.

The article of Villaseñor Herrera V., Vidales Ramos A. and Lastra J. L. M. describes the specification of a WebService-enabled Decision Support System integrated by a set of software agents. The agent-based system presented here is capable of supporting the dynamic composition and orchestration of WebServices exposed by control devices on discrete manufacturing systems

## Prospects in intelligent and distributed production control

From the existence of the previously introduced challenges and our knowledge about the current research activity in intelligent and distributed production control, it is possible to extrapolate some possible future fecund research activities in this area. To name a few:

- Bio-inspiration. Indeed, it seems to us that all distributed and intelligent approaches are led by the will to mimic nature and human behavior in their ability to self-organize and adapt to unexpected situations. Bio-inspiration could also be seen at two levels. The first one concerns the lower decision level of the system. It would be interesting to go further in that inspiration to exploit the maximum from existing solution found in the nature. Genetic algorithms, particle swarm, potential fields, bee algorithms, bat intelligence, stigmergy to name a few are typical approaches that can be deeply studied from an intelligent and distributed control point of view. The second one could be seen at a more global point of view: at the architecture level. Indeed, the studied system could be structured in the same way as a human body. The VSM (Viable System Model) proposed by S. Beer (Beer, 1984) which is especially characterize by its recursivity property, could be then an interesting way the structure and organize the agent communities. The paper of Pannequin R. and Thomas A. illustrates this prospect. Furthermore, nature provides instantiations of group-selfishness – in social insect colonies where all members are genetically related because of the queen generating all offspring – providing inspiration in decentralized design avoiding selfish routing and decision making by individual intelligent products.

- Generalization of the concept of intelligent product within its whole product life cycle, not only specifically in its production phase, which is mainly addressed by the community. The promising concept of closed-loop PLM would gain to use intelligent-based or distributed-based control system. For example, an intelligent product could enhance the exchange of information and return of experiments toward the different phases of its life cycle and the one of next generation products (from maintenance to re-design, from exploitation to production, …). Generalization could also be seen in a large amount of industrial areas or complex systems, especially in health sector or in building one, opening the way toward Product Service Systems – PSS (Morelli, 2006). In the healthcare sector, the intelligent entity concept could be applied to augment the patient with informational capabilities, for example, to the medicine subject to dispensation or the possible contraindications. In the building sector, the already known concept of BIM (Building Information Modeling) could easily be enhanced or implemented using intelligent embedded systems (eg., intelligent parts integrated in the building for safety maintenance or traceability purposes).

- The concept of distributed control can also be useful to improve the sustainability of production system, for example by embedding into products and resources energy consumption management tools. A resource should autonomously shut down its power system when required or adapts its scheduling according to new decision criteria including energy savings, which is too complex to manage from a centralized point of view. The dynamic pricing of electricity can also be more easily managed in a distributed way in production systems. The energy grid of a region can also be studied from a distributed production control point of view.

- The VIP model. When observing non-automated production, the archetypical production lines (cf. Modern Times with Charlie Chaplin) need to be redesigned for every new product model whereas the workforce needs training. In the automated version, this translates to major software maintenance for every product introduction or even non-trivial variation. This is not a feasible option in the economic sense. In contrast, very wealthy customers (VIPs) are served by a different model/organization. Every customer has an agent (butler) who searches and combines high quality services from best-in-class providers to get the customer needs fulfilled. This butler (agent) also supervises the execution of the services and handles any contingency. When in this organization a new product is introduced, no special actions are required; it is business as usual.

Therefore, this VIP organization shall be the target for future manufacturing organization, which implies the presence of intelligent products (the butlers) as well as intelligent services. It is a system design in which neither the product or the equipment decides, it is a choreography in which the interactions make the decisions emerge and in which the intelligent entities are foremost self-experts. This last property is what makes such systems cope with change.

Open systems. Our *obsession* with performance evaluation has led to the predominance of closed system designs. Indeed, the performance of an open system cannot be readily measured (analogy: try to measure travel distance performance of a navigation system that only delivers the maps but – as it is an open system component – not the routing mechanisms). Nonetheless, real progress in uncertain and complex environment will require the design and development of these open systems. In particular, the design of infrastructure reflecting the stable aspects and elements, on which to build full systems rapidly and efficiently, requires that the research community learns how to investigate this. The industrial counterpart of this ambition can be observed in reports on best practices and the needs expressed by senior personnel. For instance the MES domain, industrials prefer *visibility* (i.e. to have a kind of production radar at their disposal informing them about what will happen in the near future) over *optimization* (i.e. sophisticated systems that take the decisions based on too little or incorrect/stale information as well as limited information models).

## Acknowledgement

The guest editors acknowledge their debt to the authors and reviewers of this special issue. The guest editors address special thanks to Andrew Kusiak, Editor-in-Chief, Natalia Bakhtadze, Chair of the National Organizing Committee and Alexandre Dolgui chair of the International Program Committee for their confidence in this project.

## List of the 11 papers composing this special issue

### Other references